\documentclass{epl}

\usepackage{epsfig}
\usepackage{graphicx}

\title{Independence of the relaxation of a supercooled fluid
of its microscopic dynamics: need for yet another extension of 
the mode-coupling theory}
\author{Grzegorz Szamel \and Elijah Flenner}
\institute{Department of Chemistry, 
Colorado State University, Fort Collins, CO 80525}

\date{\today}


\begin{document}

\maketitle

\begin{abstract}
Using Brownian Dynamics computer simulations we show that the relaxation
of a supercooled Brownian system is qualitatively the same as that of
a Newtonian system. In particular, near the so-called mode-coupling transition
temperature, dynamic properties of the Brownian system exhibit the same 
deviations from power-law behavior as those of the Newtonian one. Thus, 
similar dynamical events cut off the idealized mode-coupling transition
in Brownian and Newtonian systems. We discuss implications of this finding
for extended mode-coupling theory. In addition, we point out and discuss
the difference between our findings and experimental results, 
and present an alternative interpretation of some of our simulation data. 
\end{abstract}

Our present understanding of the glass transition owes a great deal to a
series of computer simulation studies performed by Hansen and collaborators
\cite{Hansenrev}. The main conclusion from these studies is that,
at least for certain classes of materials, the glass transition is 
a two step process. The first step is
the so-called kinetic glass transition occurring at a temperature
$T_c$ above the laboratory glass transition temperature $T_g$.
The kinetic transition is associated with a sudden freezing out
of the usual hydrodynamic-like motions \cite{Hansen,HansenL}. 
This transition is usually identified with an ergodicity breaking transition
predicted by idealized mode-coupling theory (MCT)
\cite{Bengz,Mazenko,comment0}.
Below $T_c$ the dynamics is thought to be dominated by activated processes 
that are sometimes referred to as hopping events. It is claimed that 
extended MCT \cite{Das,GS,Dufty}
is able to describe these events.
               
The kinetic transition has been carefully examined in computer simulation 
studies of Andersen, Kob, and coworkers \cite{AndKob,Kob,KobNS,Kobrev}. 
The commonly accepted conclusion is that the idealized MCT
provides a good description of this transition. In particular: the
relaxation time shows a power-law temperature 
dependence with an exponent that is close to that predicted by the theory, 
the time dependence of the incoherent intermediate 
scattering function is reasonably well described by the so-called 
$\beta$-correlator of the theory, and the (effective) 
non-ergodicity parameter 
extracted from simulations agrees well with that calculated from the theory. 

The above mentioned conclusion ignores a few caveats \cite{comment}. 
One of the more fundamental ones concerns an often 
overlooked confusion about the
independence of the supercooled fluid's relaxation on its microscopic
dynamics and MCT's description of this independence.

On one hand, it is commonly expected that standard Newtonian dynamics (ND) 
leads to the same glass transition scenario as Brownian dynamics (BD). 
Indeed, the first simulational indication of the independence of
the kinetic glass transition on microscopic dynamics dates back to
Hansen's investigations \cite{HansenL}. 
At about the same time it was shown theoretically 
that the idealized MCT 
leads to the same glass transition scenario for both Newtonian
and Brownian dynamics \cite{SL}.
More recently, Gleim \textit{et al.}\ \cite{Gleim}
performed a careful comparison of simulation results obtained
using ND and so-called stochastic dynamics (SD) \cite{SD}. 
Again, they showed that the kinetic glass 
transition scenario is remarkably independent of the microscopic dynamics. 
These simulational and theoretical indications 
of the independence of the kinetic glass transition scenario 
of the fluid's microscopic dynamics justified using the MCT 
to analyze the colloidal glass transition. 
The result of this analysis was very 
favorable for the theory \cite{vanM}. 

It has been remarked \cite{vanM} that in real colloidal systems hopping
events are suppressed and thus the experimental colloidal glass transition
coincides with the idealized MCT ergodicity breaking transition.
This is good news for the theory for the 
following reason: MCT for Brownian systems cannot be extended in the same way
as MCT for Newtonian ones. In the latter case the
extension and the resulting cut-off of the ergodicity breaking transition
is achieved through coupling to current modes that are defined in terms 
of particle's velocities. Since in Brownian systems there are no velocities
(particle's positions are the only dynamic variables), there are no
currents to add to the standard set of slow variables, and the usual way
to extend MCT fails. However, 
the apparent absence of hopping events in experimental colloidal systems
made this potential problem irrelevant. Indeed, it has been
sometimes claimed that the colloidal glass transition is completely described
by the idealized MCT.

In contrast to this seemingly consistent picture, 
it was already clear from Hansen's early studies that
Brownian systems commonly investigated in computer simulations \cite{hydro} 
do not undergo the ergodicity breaking transition.
In particular, L\"owen \textit{et al.}\ \cite{HansenL} 
saw very similar hopping events for both Newtonian and
Brownian dynamics. It has to be admitted that  
these events were seen at the lowest temperature, and 
the systems studied were probably on the verge of being equilibrated. 
However, similar hopping events were seen in a later study \cite{Sood}.
In addition, 
it is clear from Gleim \textit{et al.}\ \cite{Gleim} that similar
deviations from idealized mode-coupling-like behavior are present 
in both Newtonian and stochastic systems. 

It should be emphasized that Gleim's study, although indicative, does
not really address the most important, fundamental point: qualitative
disagreement between simulations (that show absence of the ergodicity
breaking transition) and MCT for Brownian
systems (that, at present, cannot be extended). 
This is due to the fact that MCT for 
stochastic systems could, in principle, be extended in the same way as 
one for Newtonian systems (in both cases one can define current modes
using particle's velocities). It is this fact that motivated us 
to perform a large scale Brownian dynamics study of a model
used before in Newtonian and stochastic simulations: 
the Kob-Andersen \cite{AndKob} binary Lennard-Jones mixture \cite{commentLJ}.

We consider an 80:20 mixture of 1000 Lennard-Jones particles.
The interaction between the particles is given by
$V_{\alpha\beta} = 4\epsilon_{\alpha\beta}
\left[\left(\sigma_{\alpha\beta}/r\right)^{12}-
\left(\sigma_{\alpha\beta}/r\right)^{6}\right]$
where $\alpha,\beta \in \{A,B\}$ and $\epsilon_{AA}=1.0$,
$\sigma_{AA}=1.0$, $\epsilon_{AB}=1.5$, $\sigma_{AB}=0.8$,
$\epsilon_{BB}=0.5$, and $\sigma_{BB}=0.88$. All the interaction
potentials are cut-off at $2.5\sigma_{\alpha\beta}$. 
The box length of the (cubic) simulation box is equal to $9.4\sigma_{AA}$.

The Lennard-Jones binary mixture considered here was extensively
investigated using Newtonian \cite{AndKob,Kob,Kobrev} and
stochastic \cite{Gleim} dynamics. 
The main finding is that as temperature 
approaches $T_c=0.435\epsilon_{AA}/k_B$ self-diffusion coefficients
and relaxation times change rapidly; their temperature dependence
can be well fitted by power laws, in agreement with the MCT. 
However, in contrast to the
theoretical prediction, the exponents for self-diffusion coefficients 
and relaxation
times are different. Moreover, the ergodicity breaking transition 
is avoided, instead there appears to be a crossover to a different relaxation
scenario \cite{Hansenrev}. The temperature obtained from power-law fits 
is commonly referred to as the mode-coupling transition
temperature $T_c$. It should be emphasized that it is quite
different from the temperature that the MCT would
predict using realistic structure factors as input \cite{Kob}. 
Near $T_c$, time-dependence of the correlators is well described
by the MCT \cite{KobNS}; also wave-vector dependence
of the effective non-ergodicity parameter is well described by the theory.
All of these conclusions are valid for both Newtonian and stochastic dynamics.

Here we consider BD: the particles are performing 
interacting diffusive 
motion; they move under the influence of deterministic and
stochastic forces. Explicitly, for the $i$th particle,
\begin{equation}\label{Lan}
\dot{\vec{r}}_i = - \frac{1}{\xi_0} \nabla_i \sum_{j\neq i} 
V_{\alpha\beta}(|\vec{r}_i-\vec{r}_j|) + \vec{\eta}_i(t),
\end{equation}
where $\xi_0=1.0$ is the friction coefficient of an isolated particle
and random noise $\eta_i$ satisfies the fluctuation-dissipation
theorem, \textit{i.e.} 
\begin{equation}\label{FDT}
\left<\vec{\eta}_i(t) \vec{\eta}_j(t')\right> = 
2 D_0 \delta(t-t') \delta_{ij} I.
\end{equation}
In Eq.~(\ref{FDT}) $D_0=k_B T/\xi_0$ is the diffusion coefficient
of an isolated particle and $I$ is a unit tensor. One should note that 
the equations of motion allow diffusive motion
of the system's center-of-mass. All results presented here pertain
to the motion \textit{relative} to the center-of-mass.

In the following we use reduced units with $\sigma_{AA}$, $\epsilon_{AA}$,
$\epsilon_{AA}/k_B$, and $\sigma_{AA}^2\xi_0/\epsilon_{AA}$ 
being the units of length, energy, temperature, and time, respectively. 

Equations of motion (\ref{Lan}) were solved using a Heun algorithm 
\cite{Heun}. We used a very small time step of $5\times 10^{-5}$. 
To check that the equilibrium and dynamic properties are independent
of the time step, we run a couple of production runs at twice smaller 
time step. We simulated the system at temperatures 1.0, 0.8,
0.6, 0.55, 0.5, 0.47, 0.45 and 0.44. At each temperature we performed
a long equilibration run and several production runs (equilibration runs
were at least as long as production runs). The results presented were
obtained by averaging over different production runs.

One should note that at the time of the Gleim \textit{et al.}\ \cite{Gleim}
investigation the present study was not feasible. The improvement in
computer speed allowed us to perform 
very long Brownian dynamics runs (up to $6\times 10^8$ steps). 

We present here data for the mean squared displacement (MSD), 
self-diffusion coefficient, intermediate scattering function, $\alpha$ 
relaxation time, non-Gaussian parameter, and the self part of the 
van Hove correlation function. The data for the 
the first four quantities are similar to the previous results for 
Newtonian and stochastic dynamics. 
For the non-Gaussian parameter
we show for the first time that its amplitude increases
according to a power law, also the characteristic time scale seems
to increase according to a power law. The self part of the van Hove
function shows clearly that a qualitatively different relaxation scenario
sets in near $T_c$.  

\begin{figure}[t]
  \begin{minipage}[t]{0.48\textwidth}
    \begin{center}
      \includegraphics[scale=0.25]{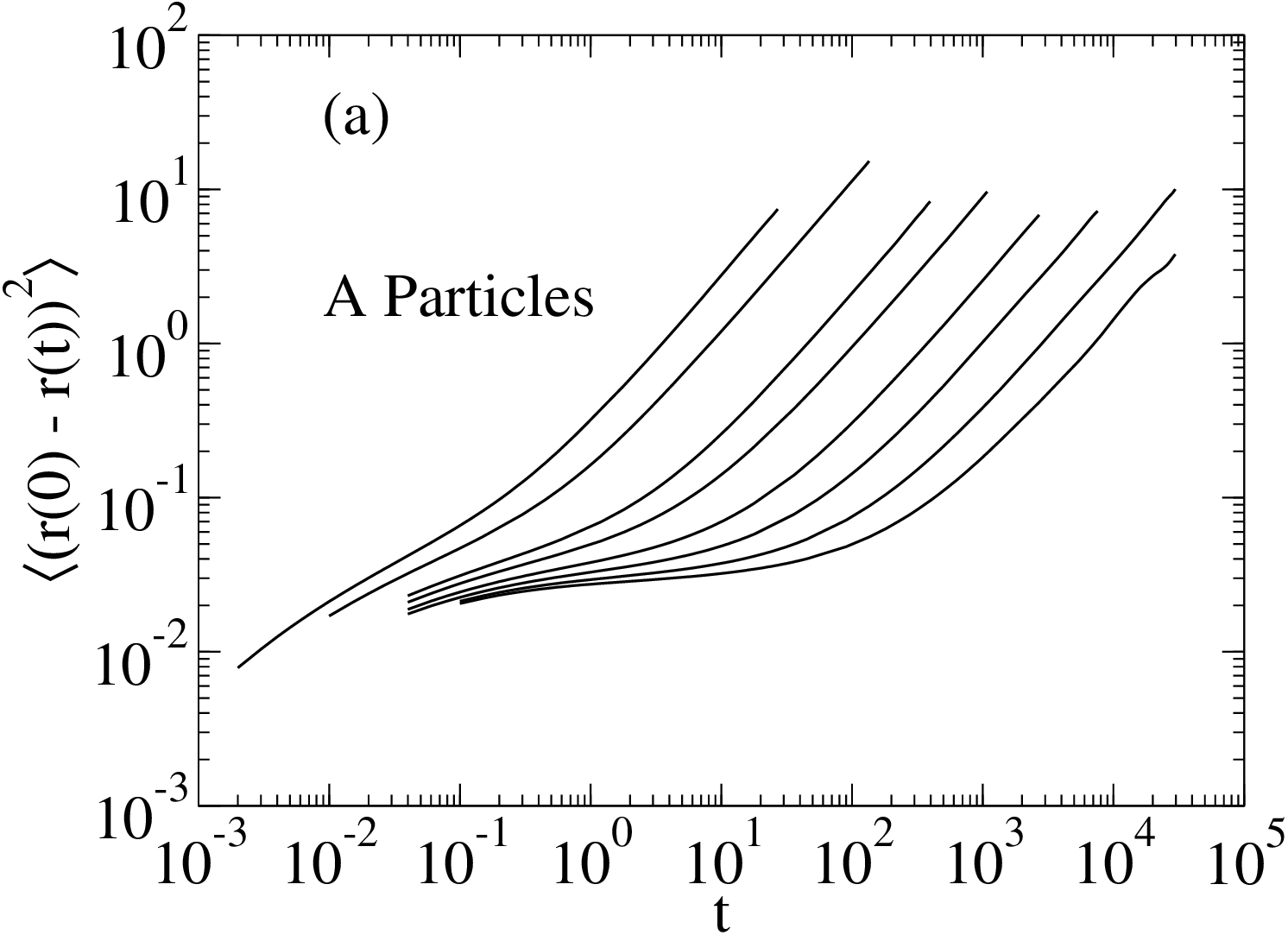}
    \end{center}
  \end{minipage}
  \hfill
  \begin{minipage}[t]{0.48\textwidth}
    \begin{center}
      \includegraphics[scale=0.25]{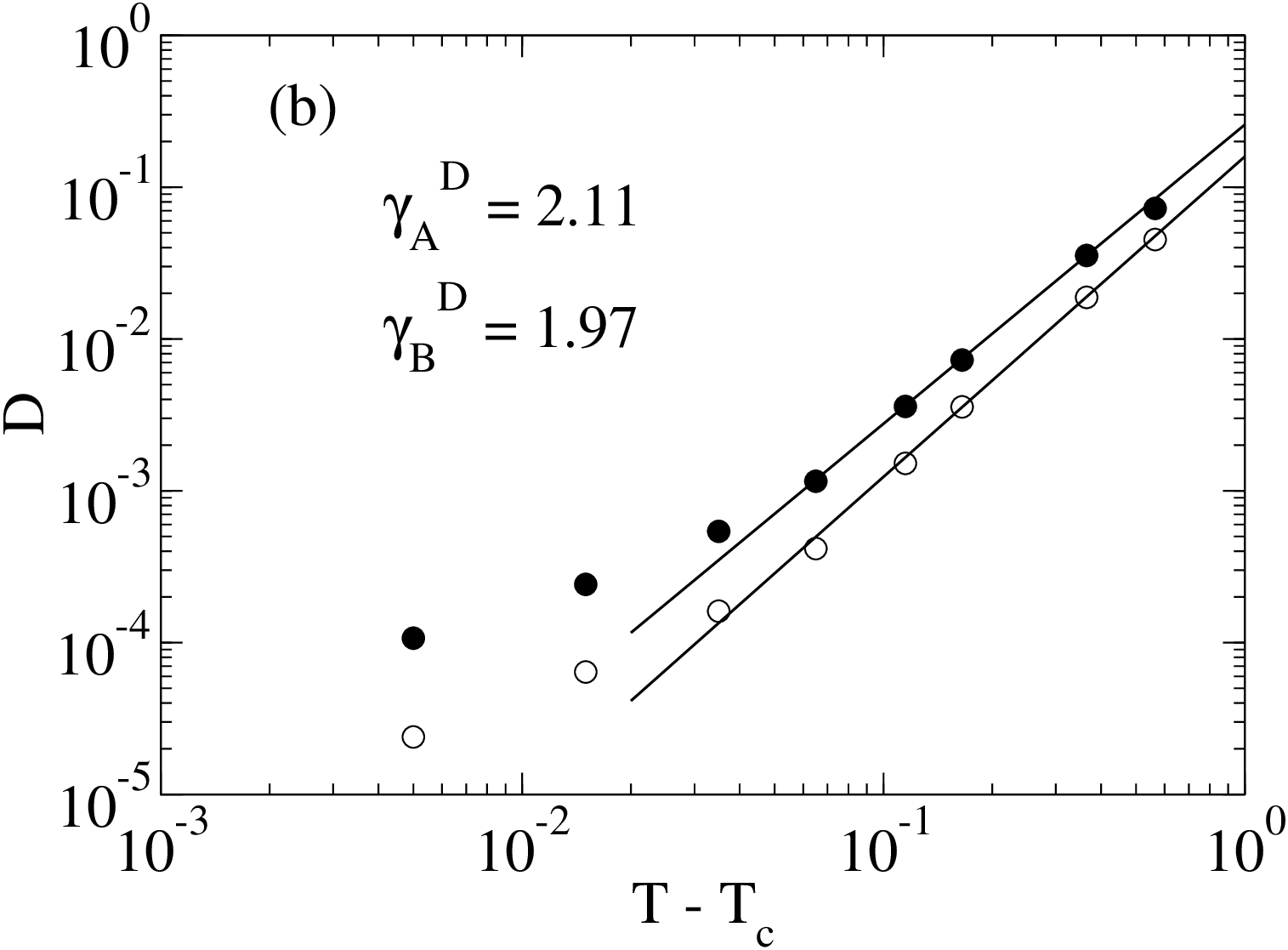}
    \end{center}
  \end{minipage}
\caption{(a) Time dependence of the mean squared displacement; 
$T=1$, 0.8, 0.6, 0.55, 0.5, 0.47, 0.45, 0.44. (b) Temperature 
dependence of the self-diffusion coefficients; open symbols: A particles, 
closed symbols: B particles; lines: power-law fits, 
$D\propto (T-T_c)^{\gamma^D}$; exponents $\gamma^D$ are listed in the figure.
\label{msddiff}}
\end{figure}

In Fig.~\ref{msddiff}a we show the MSD for the A particles.
At the shortest times the MSD 
grows linearly with time: the motion is diffusive with
diffusion coefficient equal to $D_0=k_B T/\xi_0$. In the reduced
units used here $D_0$ 
decreases by a factor of approximately 2
over the range of temperatures studied. At longer times 
the MSD changes very slowly due to transient
``cage'' localization. 
We obtain the self-diffusion 
coefficients from the diffusive behavior of the MSD
seen at the longest times. 
In Fig.~\ref{msddiff}b we show the temperature dependence of the
self-diffusion coefficients for both A and B particles: it can be fitted by 
power laws up to $(T-T_c)/T_c\approx 0.08$ 
(unless stated otherwise, all fits were performed using temperature
range $0.47\le T \le 0.8$), and the exponents are
essentially the same as in ND. Closer to $T_c$, 
the self-diffusion coefficients deviate from power laws in a way 
similar to that found in Newtonian systems.

\begin{figure}[b]
  \begin{minipage}[t]{0.48\textwidth}
    \begin{center}
      \includegraphics[scale=0.25]{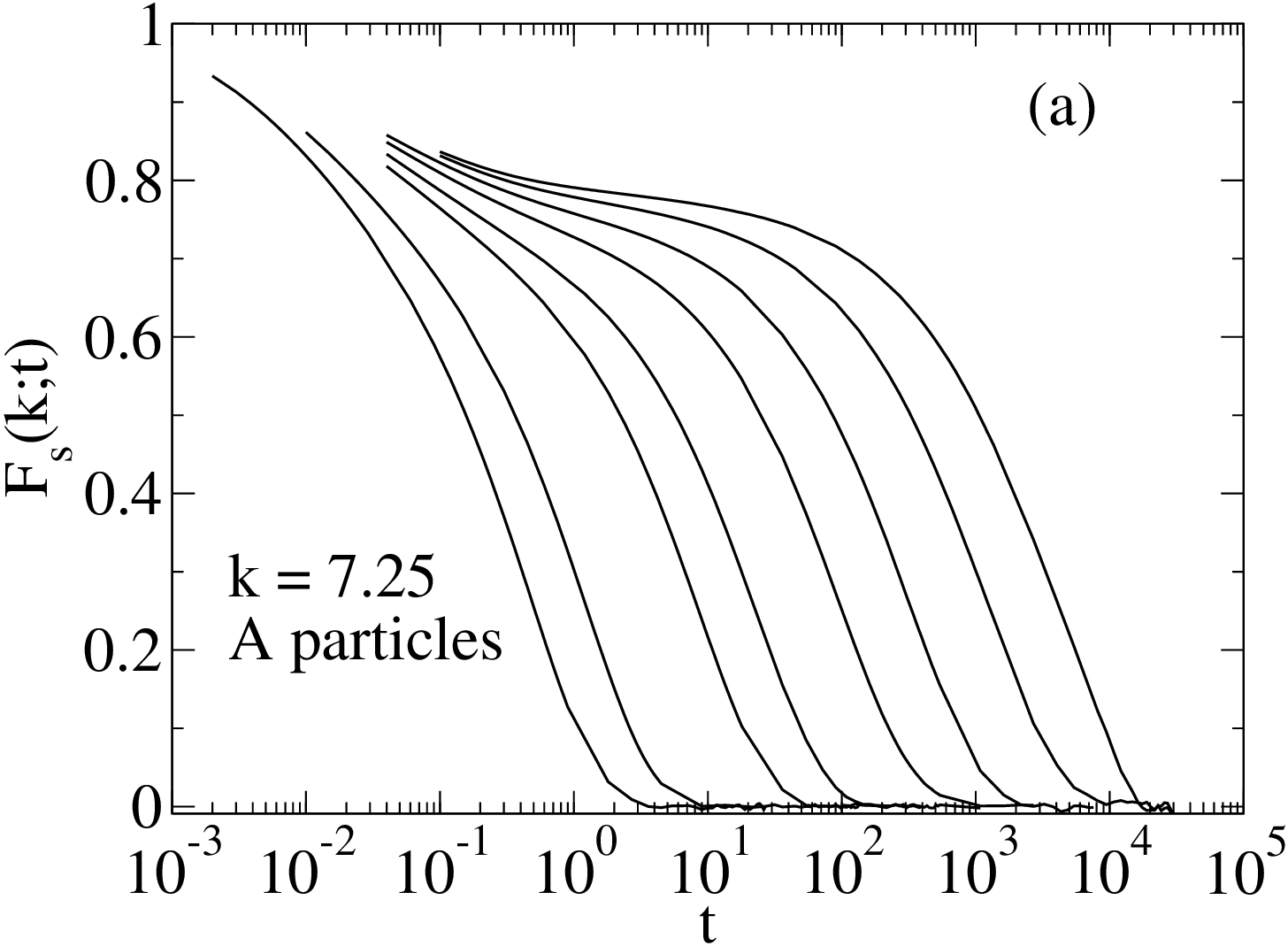}
    \end{center}
  \end{minipage}
  \hfill
  \begin{minipage}[t]{0.48\textwidth}
    \begin{center}
      \includegraphics[scale=0.25]{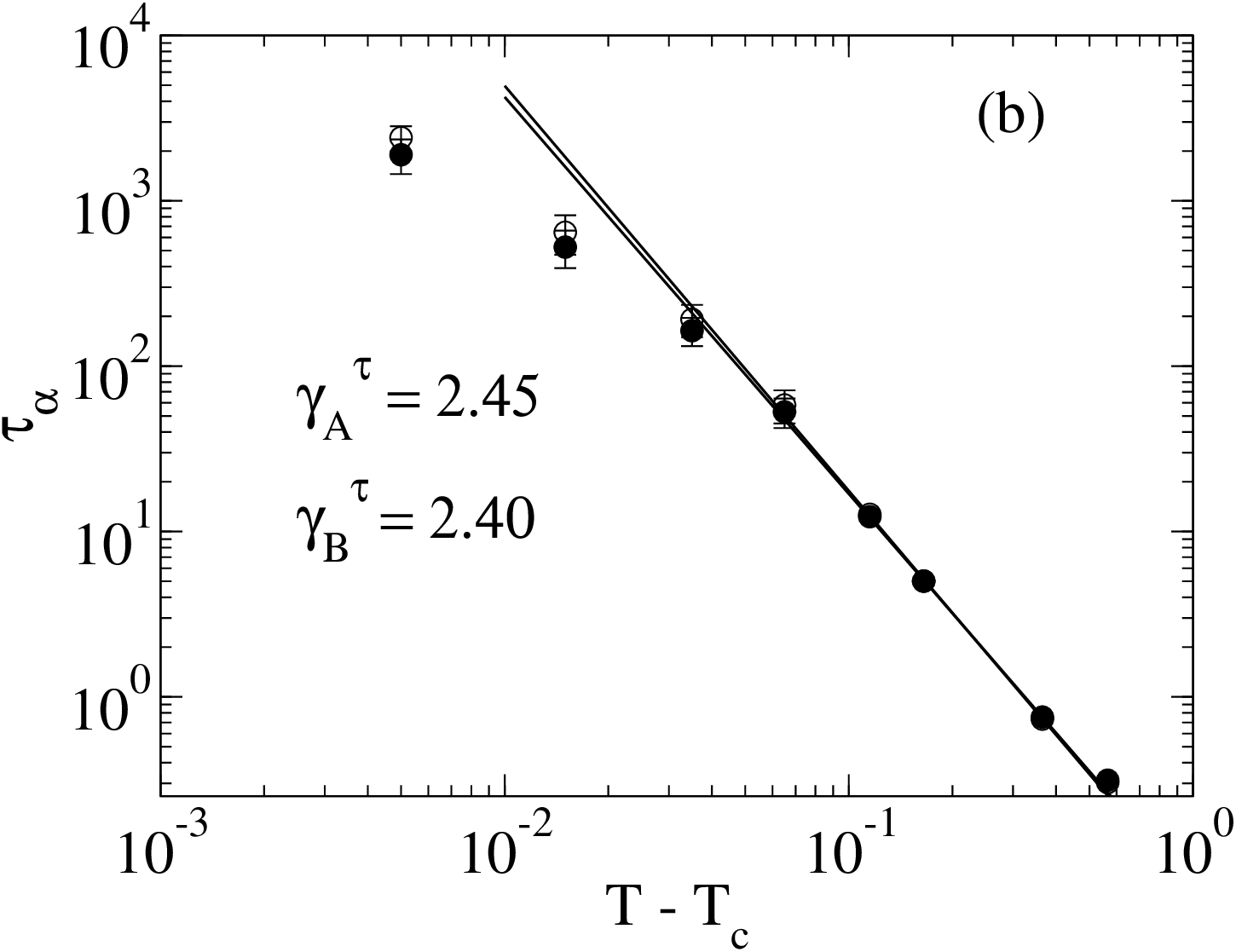}
    \end{center}
  \end{minipage}
\caption{(a) Time dependence of the incoherent intermediate scattering 
function; $T=1$, 0.8, 0.6, 0.55, 0.5, 0.47, 0.45, 0.44. (b) Temperature 
dependence of the $\alpha$ relaxation times;
open symbols: A particles, closed symbols: B particles; lines: power-law fits, 
$\tau_{\alpha}\propto (T-T_c)^{-\gamma^{\tau}}$.
\label{scattalpha}}
\end{figure}

In Fig.~\ref{scattalpha}a we show time dependence 
of  the incoherent intermediate scattering function, $F_s(k;t)$,
for all temperatures for the $A$ particles at a wavevector
close to the peak in the $AA$ partial structure factor. 
Qualitatively, we see that our results 
resemble those obtained for stochastic dynamics \cite{Gleim}:
on approaching the mode coupling transition temperature a plateau
develops while the characteristic relaxation time sharply
increases. We follow earlier workers and define the $\alpha$ relaxation
time as the time at which $F_s$ decays to $e^{-1}$. The temperature dependence
of the relaxation times for $A$ and $B$ particles is shown in 
Fig.~\ref{scattalpha}b.
Again, the temperature dependence can be fitted by power laws, and
the exponents are essentially the same as in Newtonian systems.
One should note that, as in Newtonian and stochastic systems, we
see a violation of the MCT's $\alpha$-scale 
universality: power laws for self-diffusion coefficients are
different from those for relaxation times.
As for self-diffusion coefficients, close to $T_c$ relaxation times 
clearly deviate from power laws.

\begin{figure}[t]
  \begin{minipage}[t]{0.48\textwidth}
    \begin{center}
      \includegraphics[scale=0.24]{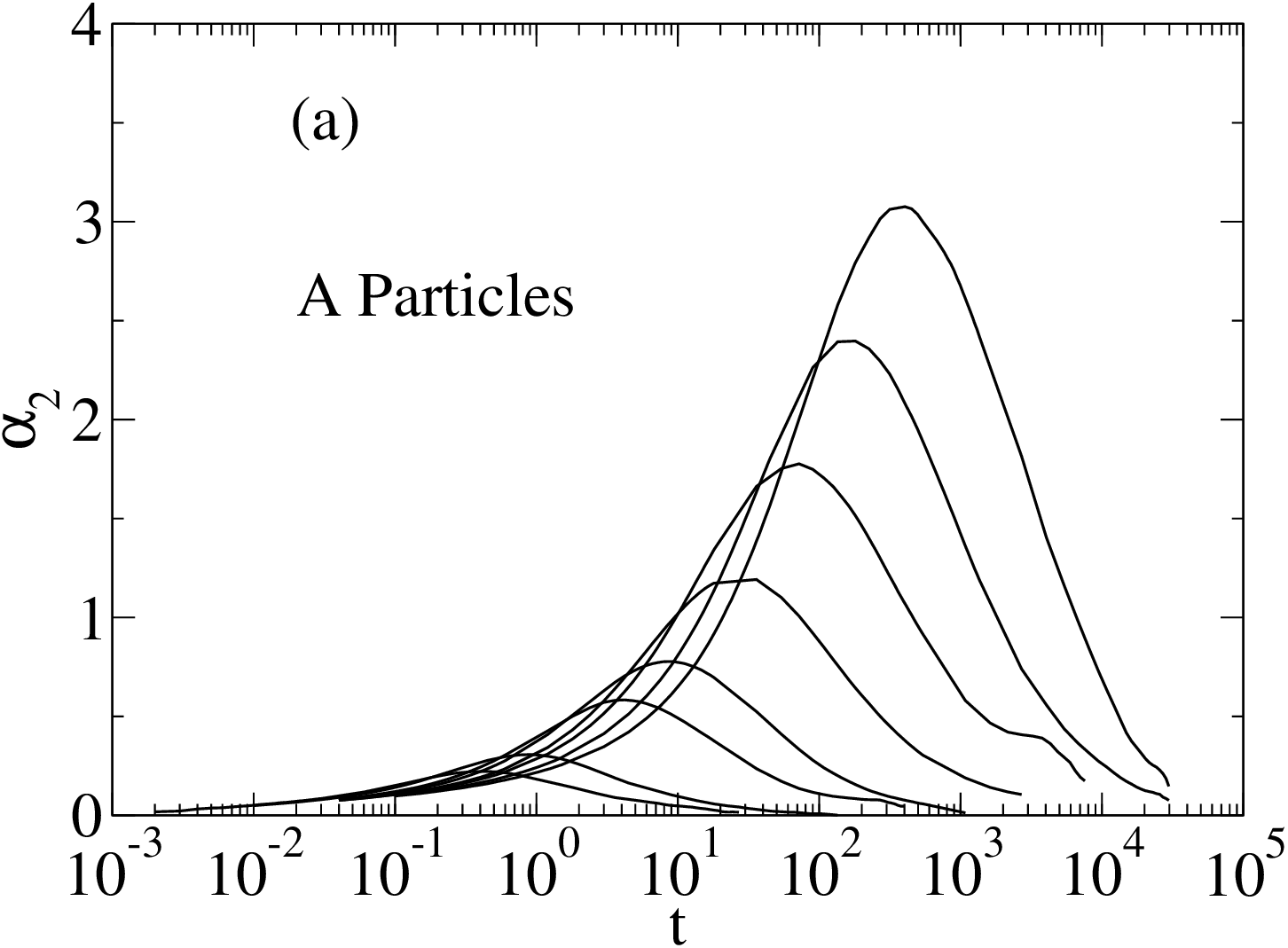}
    \end{center}
  \end{minipage}
  \hfill
  \begin{minipage}[t]{0.48\textwidth}
    \begin{center}
      \includegraphics[scale=0.24]{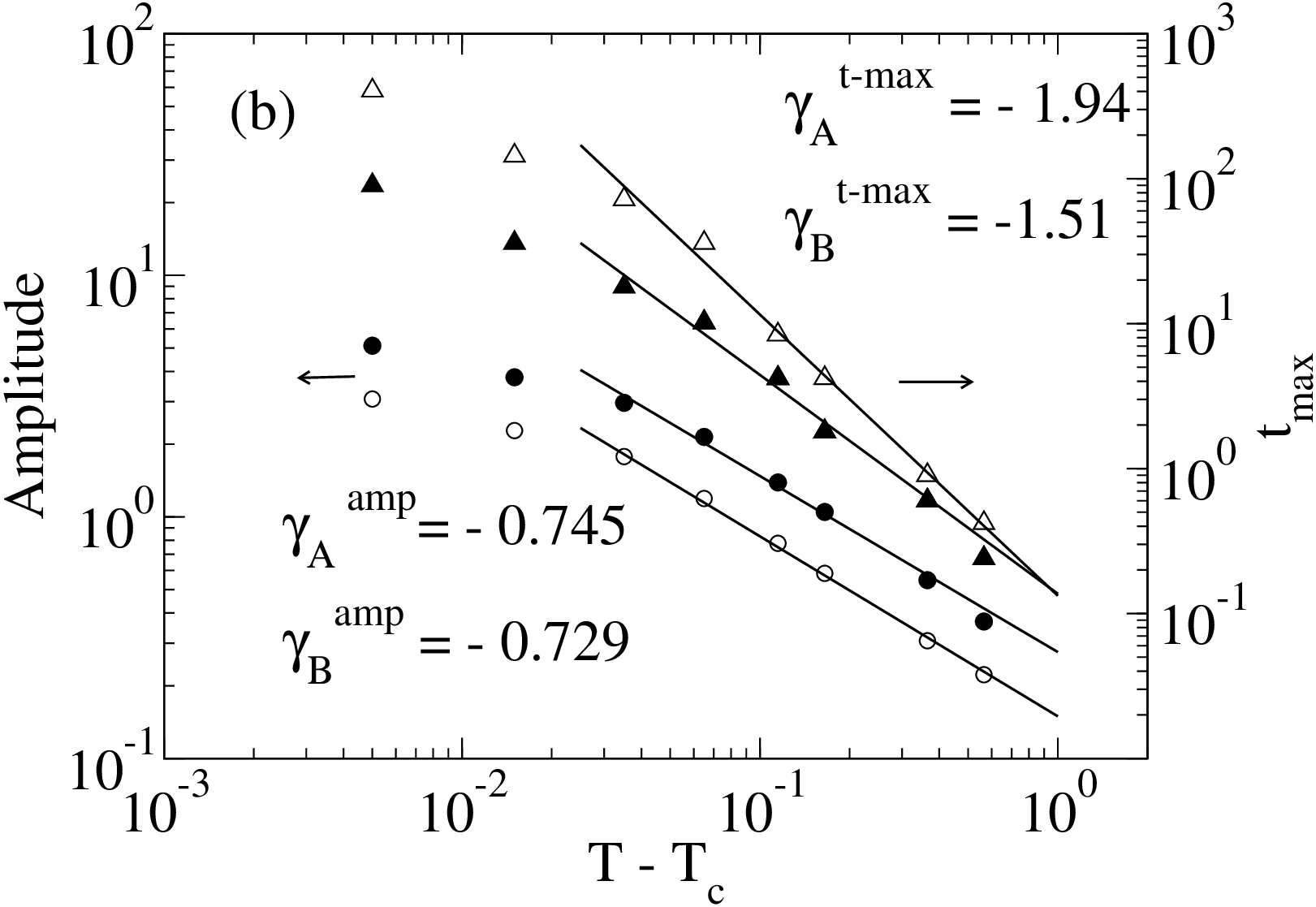}
    \end{center}
  \end{minipage}
\caption{(a) Time dependence of the non-Gaussian parameter 
function; $T=1$, 0.8, 0.6, 0.55, 0.5, 0.47, 0.45, 0.44. (b) Temperature 
dependence of the amplitude of the non-Gaussian parameter and
its characteristic time; open symbols: A particles, 
closed symbols: B particles.\label{nongauss}}
\end{figure}

In Fig.~\ref{nongauss}a we show  the non-Gaussian parameter
$\alpha_2(t) = 3\left<\Delta r^4(t)\right> / 5\left<\Delta r^2(t)\right>^2-1$.
As found earlier \cite{AndKob,Kobhet}, $\alpha_2$ 
is zero at short times, has a maximum at intermediate
times and decays to zero on time scale somewhat shorter than 
the $\alpha$ relaxation time scale.
Note that the maximum value of $\alpha_2$ is somewhat
larger than in Newtonian systems. 
In Fig.~\ref{nongauss}b we show the temperature
dependence of the maximum value of $\alpha_2$ and the time at which
it has the maximum value. Both quantities follow
power laws in the same temperature range as self-diffusion coefficients
and relaxation times. In case of the amplitude this is different 
from the prediction of the mode-coupling theory \cite{Fuchsng}: 
the theory predicts that
the amplitude of the non-Gaussian parameter saturates as temperature
approaches $T_c$. In case of the characteristic time we again see
a violation of the $\alpha$-relaxation universality: the temperature
dependence of the characteristic time of the non-Gaussian parameter
is similar to that of the self-diffusion coefficients and different
from that of the scattering function relaxation time. Finally, 
both the amplitude and the characteristic time deviate from
power law behavior close to $T_c$.

\begin{figure}[b]
  \begin{minipage}[t]{0.48\textwidth}
    \begin{center}
      \includegraphics[scale=0.24]{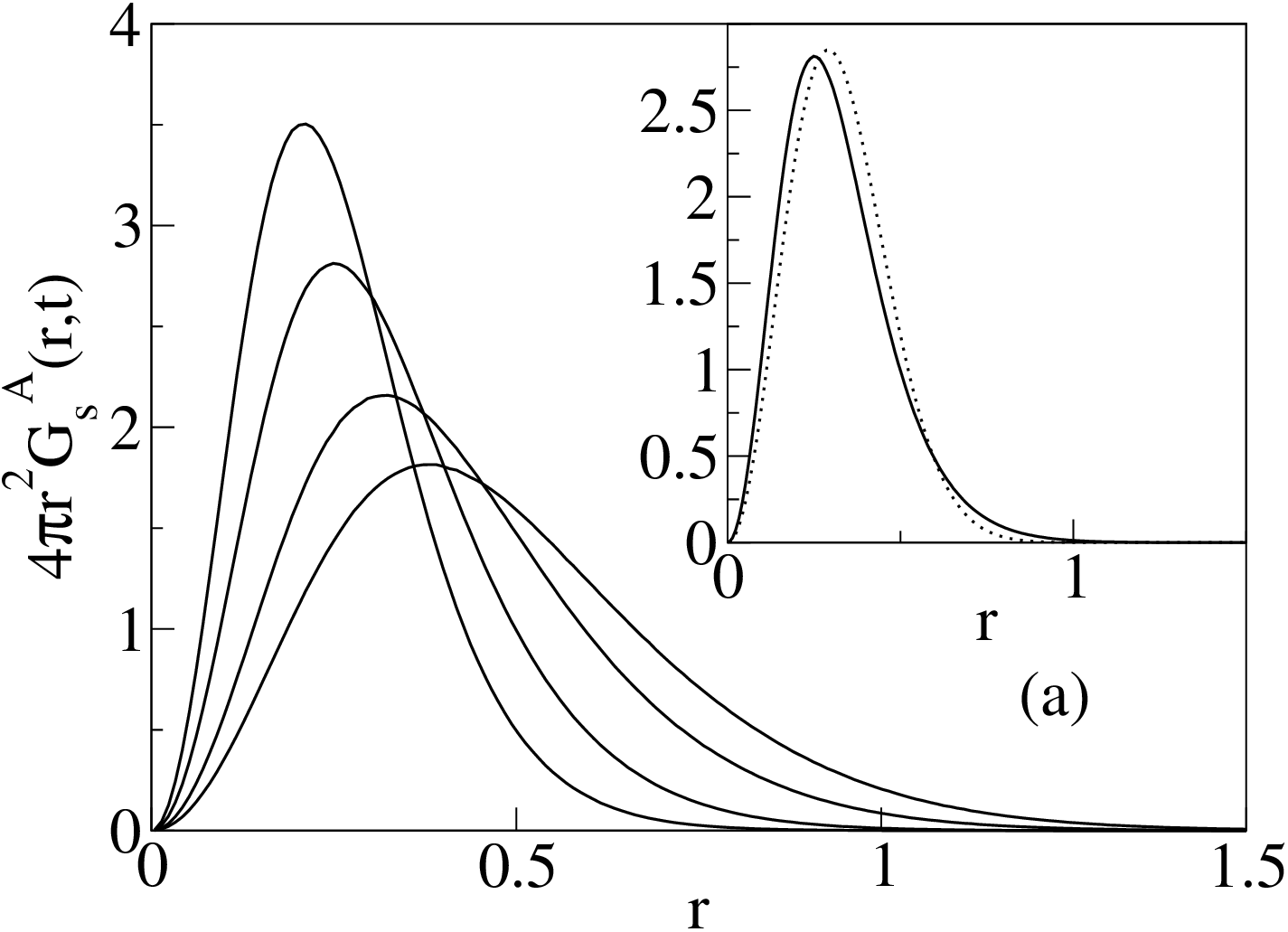}
    \end{center}
  \end{minipage}
  \hfill
  \begin{minipage}[t]{0.48\textwidth}
    \begin{center}
      \includegraphics[scale=0.24]{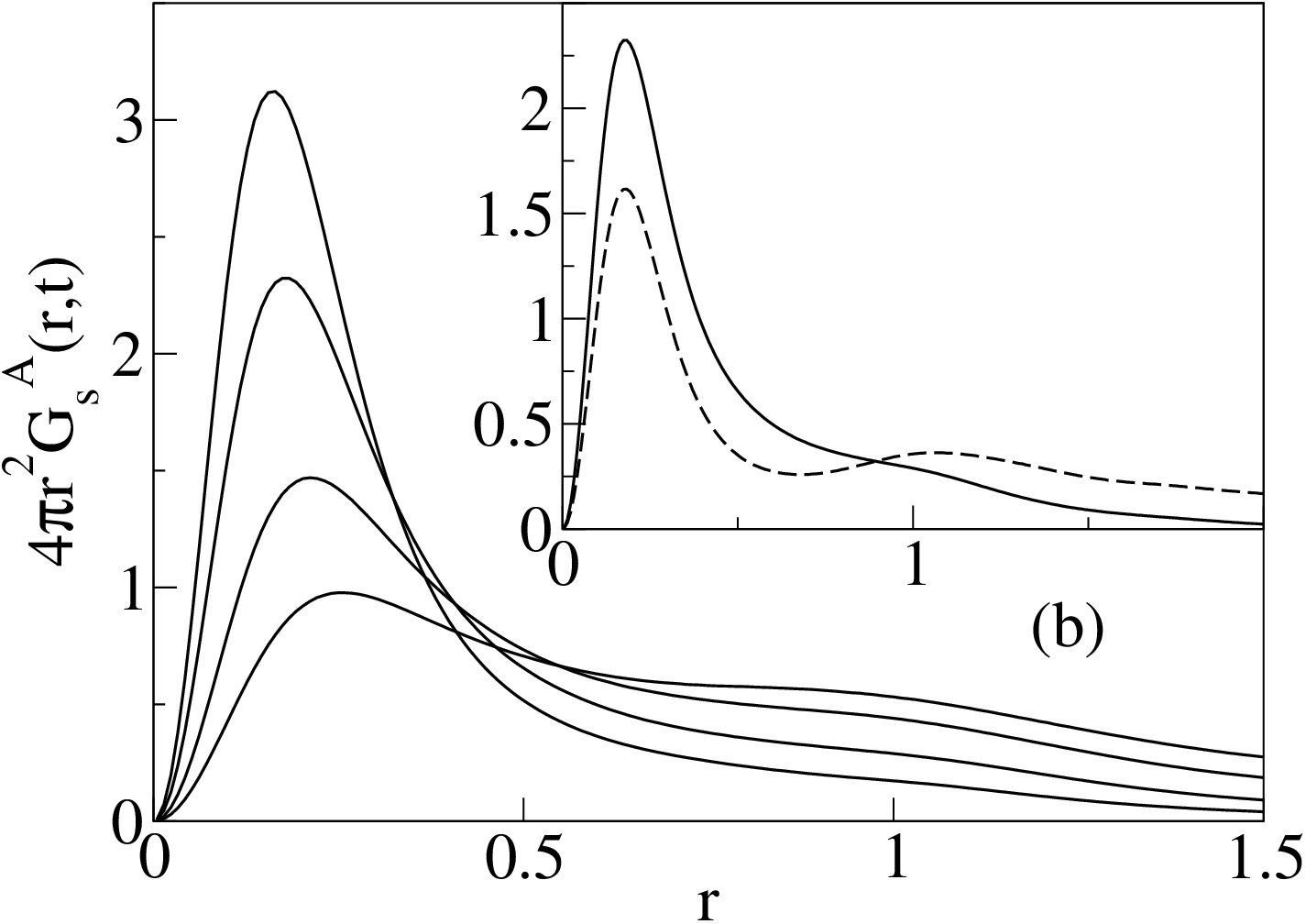}
    \end{center}
  \end{minipage}
\caption{(a) $4\pi r^2 G_s^{A}(r,t)$ at $T=1$ for $t=0.5\tau_{\alpha}$,
$\tau_{\alpha}$, $2\tau_{\alpha}$, $3\tau_{\alpha}$ 
(at $T=1$ $\tau_{\alpha} = 0.304$); insert: 
$4\pi r^2 G_s^{A}(r,t)$ at $T=1$ for $t=\tau_{\alpha}$ (solid line) 
and Gaussian probability distribution with the same $\left<r^2\right>$ 
(dotted line).
(b) $4\pi r^2 G_s^{A}(r,t)$ at $T=0.44$ for $t=0.5\tau_{\alpha}$,
$\tau_{\alpha}$, $2\tau_{\alpha}$, $3\tau_{\alpha}$ 
(at $T=0.44$ $\tau_{\alpha} = 2409$); insert: 
$4\pi r^2 G_s^{A}(r,t)$ (solid line) and $4\pi r^2 G_s^{B}(r,t)$ (dashed
line) at $T=0.44$ for $t=\tau_{\alpha}$.\label{vanHove}}
\end{figure}

The last quantity presented here, shown in Fig.~\ref{vanHove}, is the
self part of the van Hove correlation function, 
$G_s^{\alpha}(r,t) = \frac{1}{N_{\alpha}}
\left<\sum_{i=1}^{N_{\alpha}}\delta\left(r-|\vec{r}_i(t)-\vec{r}_i(0)|\right)
\right>,$
where $\alpha \in \{A,B\}$. We show $4\pi r^2 G_s^A(r,t)$
for times equal to 
$0.5\tau_{\alpha}$, $\tau_{\alpha}$, $2\tau_{\alpha}$ and $3\tau_{\alpha}$
at two different temperatures, $T=1$ (Fig.~\ref{vanHove}a) and
$T=0.44$ (Fig.~\ref{vanHove}b). We clearly see two different relaxation
scenarios. At the higher temperature 
the position of the peak of the van Hove function moves with time and
the shape of the correlation function is close to Gaussian.
At the lower temperature the position of the peak is almost
time-independent. In addition, a pronounced shoulder develops at 
distances around 1. In the same time range  
$4\pi r^2 G_s^{B}$ 
has a well developed peak at distances around 1 (see insert). 
Similar features in the van Hove function were seen earlier
\cite{Hansen, HansenL} but only at the lowest temperature.
In this study an arrested peak of the van Hove function, a shoulder in
$4\pi r^2 G_s^{A}$, and a secondary peak in $4\pi r^2 G_s^{B}$ were
seen both at $T=0.44$ and $T=0.45$ (not shown). 
Earlier studies 
\cite{Hansen, HansenL} interpreted such features 
as evidence of an activated hopping process.
Note that at $T=0.44$ for the longest time shown in 
Fig.~\ref{vanHove}b, $t=3\tau_{\alpha}$,
the van Hove function is very clearly far from Gaussian although 
the non-Gaussian parameter for $t=3\tau_{\alpha}$ is well past its 
maximum. The non-Gaussian shape of the van Hove function 
suggests that pronounced dynamic heterogeneity is still
present on this time scale. 

In summary, we clearly showed that in Brownian systems,
as found earlier in Newtonian and stochastic systems, 
the kinetic glass transition does not coincide with the ergodicity
breaking. The power law dependence of transport properties,
predicted by the mode-coupling theory, is valid over two to three
decades of change of the self-diffusion coefficients and
relaxation times. Close to $T_c$  there
are departures from the power laws that cannot be (even qualitatively) 
explained by the mode-coupling theory.
The deviations from power laws are similar to those found for Newtonian
and stochastic systems \cite{Gleimth}. This confirms an earlier observation
\cite{HansenL} that similar dynamic events are responsible for relaxation
in Newtonian and Brownian systems 
below the kinetic glass transition temperature. 

The results presented here suggest a need for an extension of the
idealized mode-coupling theory that is applicable to both 
Newtonian and Brownian systems. At present it is unclear how 
to proceed toward this goal. It should be noted in this context 
that a mode-coupling-like theory for kinetic Ising models with 
stochastic dynamics has been extended \cite{Andersen}. The final
formulas of the extended theory resemble those of the extended
mode-coupling theories \cite{GS,Das,Dufty} although they are derived
in a completely different way.  

We would like to point out the disagreement between our
findings and the experimental results. The most striking difference
is between our Figs.~\ref{msddiff}b and \ref{scattalpha}b and Fig.~11 
of van Megen \textit{et al.}\ \cite{vanM2}:
our results violate the $\alpha$-scale universality and deviate
from power laws close to the transition temperature whereas 
van Megen's results show the same power laws for the self-diffusion
coefficients and the relaxation time and do not show any deviation
from power-law behavior \cite{commentshort}.
There could be two possible sources of this disagreement. First,
our particles interact via Lennard-Jones potentials whereas  
van Megen's colloidal particles have hard-sphere-like interactions. 
Second, we did not include hydrodynamic interactions that are
known to be very important in real hard-sphere suspensions.
Instead of speculating on the importance of these two issues
we would like to point out that \textit{neglecting hydrodynamic interactions}
it should be possible to simulate
a polydisperse hard sphere system that quantitatively models the
experimental system of Ref.~\cite{vanM2}. 
Such a simulation could exclude the difference in 
the interaction potentials as the source of the qualitative disagreement 
between simulations and experiment. 

Finally, we would like to note that an alternative interpretation \cite{thanks}
of \emph{some} of our simulation data is possible: 
if the transition temperature is a fit parameter,
reasonable power-law fits 
can be obtained for self-diffusion coefficients and relaxation times
in the low temperature region, $0.44\le T\le 0.6$. These
fits result in a considerably lower transition temperature 
$T_c' \approx 0.414$ ($T_c'$ depends
slightly on the quantity fitted) and the exponents 
$\gamma_A^D=2.56$, $\gamma_B^D=2.09$, 
$\gamma_A^{\tau}=3.04$, and $\gamma_B^{\tau}=2.93$.
While these fits suggest that the power-law temperature dependence of
the self-diffusion coefficients and relaxation times may extend
to temperatures lower than reported previously \cite{AndKob,Gleim}, 
accepting this alternative
interpretation would open a new set of questions. Here we pose only 
two of them: 
First, does the kinetic
transition temperature depend on the microscopic dynamics in 
stark disagreement with the prediction of the idealized MCT \cite{SL},
or do the previously reported data for ND \cite{AndKob} 
and SD \cite{Gleim} have to be re-interpreted as well? Second, how to reconcile
the validity of the idealized MCT-like power-law temperature dependence 
and the strikingly non-Gaussian van Hove function
that suggests presence of hopping-like processes that are absent in the 
idealized MCT?

\bigskip
ACKNOWLEDGMENT
\medskip
                           
Support by NSF Grant No. CHE 0111152 is gratefully acknowledged.


\begin{thebibliography}{99}
\bibitem{Hansenrev} J.P. Hansen, 
Physica A \textbf{201}, 138 (1995).
\bibitem{Hansen} J.N. Roux, J.L. Barrat, and J.P. Hansen, 
J. Phys. Cond. Matt. \textbf{1}, 7171 (1989); 
J.L. Barrat, J.N. Roux, and J.P. Hansen, 
Chem. Phys. \textbf{149}, 197 (1990).
\bibitem{HansenL} H. L\"owen, J.P. Hansen, and J.N. Roux, 
Phys. Rev. A \textbf{44}, 1169 (1991).
\bibitem{Bengz}U. Bengzelius, W. G\"otze, and A. Sj\"olander, 
J. Phys. C \textbf{17}, 5915 (1984).
\bibitem{Mazenko}S.P. Das, G.F. Mazenko, S. Ramaswamy, and J.J. Toner,
Phys. Rev. Lett. \textbf{54}, 118 (1985).
\bibitem{comment0} To distinguish between the original 
MCT \cite{Bengz,Mazenko} and its later, extended versions
\cite{Das,GS,Dufty}, whenever ambiguity can arise, 
we will refer to the former one as ``the idealized mode-coupling
theory''. 
\bibitem{Das} S.P. Das and G.F. Mazenko, Phys. Rev. A \textbf{34}, 2265 (1986).
\bibitem{GS} W. G\"otze and L. Sj\"ogren, Z. Phys. B \textbf{65}, 415 (1987).
\bibitem{Dufty} 
R. Schmitz, J. W. Dufty, and P. De, 
Phys. Rev. Lett. \textbf{71}, 2066 (1993).
\bibitem{AndKob} W. Kob and H.C. Andersen, Phys. Rev. E \textbf{51}, 
4626 (1995); 
Phys. Rev. E \textbf{52}, 4134 (1995).
\bibitem{Kob} M. Nauroth and W. Kob, Phys. Rev. E \textbf{55}, 657 (1997).
\bibitem{KobNS} W. Kob, M. Nauroth, and F. Sciortino, J. Non-Cryst. Solids
\textbf{307}-\textbf{310}, 181 (2002).
\bibitem{Kobrev} For a review see, {\it e.g.}, W. Kob,
J. Phys. Cond. Matter \textbf{11}, R85 (1999).
\bibitem{comment} The most often mentioned one concerns the location 
of the kinetic transition: when computer simulated structure factors
(that constitute the only input required by the theory)
are used in mode-coupling calculations the resulting transition
temperature for the Lennard-Jones system studied by Kob \textit{et al.}\ is
overestimated by a factor of two \cite{Kob}. 
\bibitem{SL} G. Szamel and H. L\"owen, Phys. Rev. A \textbf{44}, 8215 (1991).
\bibitem{Gleim} T. Gleim, W. Kob, and K. Binder, 
Phys. Rev. Lett. \textbf{81}, 4404 (1998).
\bibitem{SD} Stochastic dynamics is a Newtonian dynamics 
with addition of a Gaussian distributed
white noise force and a damping force proportional to the particle's
velocity.
\bibitem{Sood} S. Sanyal and A.K. Sood, 
Phys. Rev. E \textbf{57}, 908 (1998).
\bibitem{vanM} W. van Megen, Transport Theory and Stat. Phys. \textbf{24},
1017 (1995).
\bibitem{hydro} Note that in computer simulations of colloidal suspensions
hydrodynamic interactions are usually neglected.
\bibitem{commentLJ} We are using the Kob-Andersen mixture rather than 
a more realistic screened-Coulomb (\textit{i.e.}\ Yukawa) system
in order to be able to directly compare our Brownian Dynamics results
to those obtained for the same system using Newtonian \cite{AndKob}
and stochastic \cite{Gleim} dynamics.
\bibitem{Heun} W. Paul and D.Y. Yoon, Phys. Rev. E \textbf{52}, 657 (1995).
\bibitem{Kobhet} W. Kob \textit{et al}, Phys. Rev. Lett. \textbf{79} 2827 
(1997).
\bibitem{Fuchsng} M Fuchs, W. G\"otze, M.R. Mayr,
Phys. Rev. E \textbf{58}, 3384 (1998).
\bibitem{Gleimth} T. Gleim, PhD Dissertation, unpublished.
\bibitem{Andersen} S.J. Pitts and H.C. Andersen, 
J. Chem. Phys. {\bf 114}, 1101 (2001).
\bibitem{vanM2} W. van Megen, T.C. Mortensen, S.R. Williams and
J. M\"uller, Phys. Rev. E \textbf{58}, 6073 (1998).
\bibitem{commentshort} Note that simulational 
results vary over approximately three decades whereas the 
experimental ones vary over approximately four decades. However, if 
long-time diffusion coefficients are normalized by the short-time
ones, in both cases the results vary over about three decades.
\bibitem{thanks} We thank an anonymous referee for the suggestion.
\end{thebibliography}
\end{document}